# Cluster Analysis for a Scale-Free Folksodriven Structure Network

Massimiliano Dal Mas

me @ maxdalmas.com

*Abstract* — Folksonomy is said to provide a democratic tagging system that reflects the opinions of the general public, but it is not a classification system and it is hard to make sense of. It would be necessary to share a representation of contexts by all the users to develop a social and collaborative matching. The solution could be to help the users to choose proper tags thanks to a dynamical driven system of folksonomy that could evolve during the time. This paper uses a cluster analysis to measure a new concept of a structure called "*Folksodriven*", which consists of tags, source and time. Many approaches include in their goals the use of folksonomy that could evolve during time to evaluate characteristics. This paper describes an alternative where the goal is to develop a weighted network of tags where link strengths are based on the frequencies of tag co-occurrence, and studied the weight distributions and connectivity correlations among nodes in this network. A practical example is used to demonstrate the power of the Folksodriven concept and its application in practice. The paper proposes and analyzes the network structure of the Folksodriven tags thought as folksonomy tags suggestions for the user on a dataset built on chosen websites. It is observed that the hypergraphs of the Folksodriven are highly connected and that the relative path lengths are relatively low, facilitating thus the serendipitous discovery of interesting contents for the users. Then its characteristics, Clustering Coefficient, is compared with random networks. A discussion about the observed differences is provided. The goal of this paper is a useful analysis of the use of folksonomies on some well known and extensive web sites with real user involvement. The advantages of the new tagging method using folksonomy are on a new interesting method to be employed by a knowledge management system.

## I. Introduction

Nowadays the social conventions and social context is recreated through the Web 2.0, defined as a second generation, or more personalized, communicative form of the World Wide Web that emphasizes the "Social Web" through: active participation, connectivity, collaboration and sharing of knowledge and ideas among users.

Web 2.0 applications include user-centric publishing and knowledge management platforms like Wikis, Blogs, and social web resource sharing systems using social and collaborative matching. In the context of knowledge sharing the Ontology defines a common set-of-concept, but unfortunately ontologies are not wide spread at the moment. Instead Folksonomy is said to provide a democratic tagging system that reflects the opinions of the general public, but it is not a classification system and it is difficult to make sense of [1, 2]. The term "folksonomy" indicates the emergent labeling of lots of resources by people in a social context, typically flat name-spaces. The emergent data from the actions of millions of ordinary untrained "folks" can be used by different kind of tools that immediate can benefit for each individual user. The context of the use in these systems is not just one of personal organization, but of communication and sharing between people sharing their own local interests. The near instant feedback in these systems leads to a communicative nature of tag use.

It would be necessary to share a representation of contexts by all the users. The solution could be to help the users to choose proper tags thanks to a dynamical driven system of folksonomy that could evolve during the time. In this paper we identify syntactical phrases in natural language sentences to analyze the main network characteristics of articles from a chosen group of websites. After having extracted their underlying data structures towards folksonomy tags, we correlate those with the source and its time exposition, as a measure of the time of its disposal. Considering those we define a structure called *"Folksodriven"*, and adapt classical network measures to them. Then its Clustering Coefficient is compared with random networks insight on the reported network metric. A discussion about the observed differences is provided.

Folksodriven notation can be used in a community environment to make explicit groups of interests which already exist intrinsically in a Folksodriven system, to be used for recommender systems, opinion mining, and sentiment analysis.

Possible applications for such framework based on that new method can be employed by different kinds of knowledge management systems that needs to use the knowledge sharing by different people like Wikis, Blogs and all the "social web" applications, and could be used on personalized advertising system.

The paper is structured as follows:
- Section 2 introduces the notation and the necessary background;
- Section 3 introduces the Folksodriven notation;
- Section 4 extracts the underlying data structures towards folksonomy tags for the datasets which our analyses is based on;

---

- Section 5 introduces the Clustering Coefficient to measures the network properties;
- Section 6 examines a Folksodriven graph by studying the structure of the tags' network;
- Section 7 illustrates the related work;
- Section 8 summarizes some conclusions, and finally
- Section 9 proposes future work and open issues on the topic illustrated.

## II. FOLKSONOMY AND THE SEMANTIC WEB

This section introduces the notation and the necessary background for this article. It first introduces the folksonomy tag and then the correlation between the folksonomy and the semantic web.

### A. Folksonomy tag

Folksonomies are concentrations of user-generated categorization principles. The term is a neologism consisting of a combination of the words folk and taxonomy. Taxonomy is from the Greek taxis and nomos. Taxis means classification and nomos means management. Literally, it may be translated to "people's classification management".

Users can freely choose tags creating a folksonomy. Folksonomy tags may be said to be metadata from/for the masses [1]. Folksonomy tags are thus created by the people for the people on the basis of the premise that the categorizing people can create a categorization that will better reflect the people's conceptual model, contextualization and actual use of the data. The distinctive feature of folksonomies is that it is not a classification in a strict sense, but they consist of disconnected and loosely related keywords only connected by associative relations. Here, there is no hierarchy between superior and subordinated concepts [2].

### B. Ontology and Folksonomy

An ontology is a formal specification of a conceptualization of an abstract representation of the world or domain we want to model for a certain purpose.

Ontologies can capture the semantics of a set of terms used by some community: but meanings change over time, and are based on individual experiences, and logical axioms can only partially reflect them. Web developers encodes all the information in an ontology filled with rules that say, essentially, that "Robert" and "Bob" are the same. But humans are constantly revising and extending their vocabularies, for instance at one times a tool might know that "Bob" is a nickname for "Robert," but it might not know that some people named "Robert" use "Roby", unless it is told explicitly [3].

There is a perception of ontologies as top-down authoritarian constructs, not always related to the variety resources in a domain context of the ontologies. So it is understandable that bottom-up structures like folksonomy seams more attractive. Folksonomy is what has emerged as the de facto description of online content, used by million sites already worldwide to describe every kind of uri resource. Folksonomy works as the semantic web even if it is not the official semantic web, not following a unique standard or organization.

### C. Networks concepts and Folksonomy tags

Network concepts are also known as network statistics or network indices. Examples: connectivity (degree), clustering coefficient, topological overlap, etc.

Network concepts underlie network language and systems modeling. Dozens of potentially useful network concepts are known from graph theory.

## III. NOTATION USED

In this section, it is introduced the formal notation used in the paper for the Folksodriven and the related Time exposition.

In a model of space-time, every point in space has four coordinates (x, y, z, t), three of which represent a point in space, and the fourth a precise moment in time. Intuitively, each point represents an event, an event that happened at a particular place at a precise moment. The usage of the four-vector name assumes that its components refer to a "standard basis" on a Minkowski space [4].

Points in a Minkowski space are regarded as events in space-time. On a direction of time for the time vector we have:
- past directed time vector, whose first component is negative, to model the "history events" on the Folksodriven notation
- future directed time vector, whose first component is positive, to model the "future events" on the Folksodriven notation.

### A. Folksodriven Notation

$$(1) \quad FD := (C, E, R, X)$$

A folksodriven will be considered as a tuple (1) defined by finite sets composed by the *Formal Context* (*C*), the *Time Exposition* (*E*), the *Resource* (R) and the ternary relation X [5]. A *Formal Context* (*C*) is a triple $C := (T, D, I)$ where the objects *T* and the attributes *D* are sets of data and *I* is a relation between *T* and *D* [6] – see *IV.B*.

For the *Time Exposition* (*E*) we chose the Clickthrough rate (CTR) that is the number of clicks on a *Resource* (*R*) divided by the number of times that the *Resource* (*R*) is displayed (impressions).

The *Resource* (*R*) is represented by the uri of the webpage that the user wants to correlate to a chosen tag.

While *X* is defined by the ternary relation $X = C \times E \times R$ in a vector space delimited by the vectors *C*, *E* and *R*.

Following the above notation an article *A* can be depicted as a relation (2) on the Folksodriven sets.

$$(2) \quad A(c,r) := \{(c,e,r) \in X \mid e \in E\}$$

## IV. FOLKSODRIVEN DATA SET

In this section we will introduce the data set, and how it was built. The data set has been built from articles taken during a month from the: Wall Street Journal, New York Times and Financial Times web sites. Tokens were extracted from the title and description of the articles. Those tokens compose a data set of words proposed to the users as tags that he/she can add to a document - the articles on the web sites - to describe it. The near instant feedback in this system leads to a communicative nature of tag use. We see how the "Folksodriven Data Set" can "drive" the user on the choice of a correct folksonomy tag.

### A. Shallow parsing for folksonomy

Shallow Parsing is the process of identifying syntactical phrases in natural language sentences. Identifying whole parse trees can provide deeper analyses of the sentences respect identifying the chunks, but chunking is a less hard problem.

Chunking is a bottom-up unordered flat set of keywords, but it's not a classification system.

A chunking operation segments text into an unstructured sequence of text units called "chunks" [7]. A chunk is a textual unit representing a fragment of information, it consists of a single content word (a head), with the possible addition of some function or modifiers words [8].

For instance a sentence on the title element of an article

```
Stocks seesaw in volatile trade
```

will be chunked by flattening down the parse tree identifying non-overlapping and non-embedded phrases as follows:

```
[NN Stocks] [NN seesaw] [PP in] [NP volatile trade]

1.  [Stocks]    Noun Phrase
2.  [seesaw]    Noun Phrase
3.  [in]        Preposition Phrase
4.  [volatile]  Noun Phrase
5.  [trade]     Noun Phrase
```

The sentence is segmented into five chunks each including a sequence of adjacent word tokens mutually related through dependency links. The output of chunking is under specification having omitted the features for the intra-chunk, and it is non-exhaustive: some words in a sentence may not be grouped into a chunk. Among the determined chunks only the Noun Phrases and Verb Phrases were considered for tokenization using stop words for adjectives and adverbs, prepositions and numbers.

```
|Stocks | seesaw| volatile| trade|
```

From the article title the above words were considered.

The same procedure was followed for the description element. Only the Noun Phrases and Verb Phrases were considered:

```
1.  [U.S.]              14. [investors]
    Noun Phrase             Verb Phrase
2.  [stocks]            15. [hesitant]
    Noun Phrase             Noun Phrase
3.  [slid]              16. [to]
    Verb Phrase             Prep. phrase
4.  [in]                17. [make]
    Prep. Phrase            Verb phrase
5.  [choppy]            18. [big]
    Noun Phrase             Noun phrase
6.  [trading]           19. [bets]
    Verb Phrase             Noun phrase
7.  [as]                20. [amid]
    Prep. Phrase            Prep. Phrase
8.  [a]                 21. [worries]
    Prep. Phrase            Noun phrase
9.  [dearth]            22. [about]
    Noun Phrase             Noun phrase
10. [of]                23. [a]
    Prep. Phrase            Prep. phrase
11. [economic]          24. [potential]
    Noun Phrase             Noun phrase
12. [news]              25. [global]
    Noun Phrase             Noun phrase
13. [left]              26. [recesion]
    Noun Phrase             Noun phrase
```

Numbers and prepositions were considered as stop words.

### B. Incidence of context on folksonomy technology

Throughout this paper we will use the notion context in the sense of formal context as used in the ontological sense as well in Formal Concept Analysis (FCA), a branch of Applied Mathematics.

FCA is a method mainly used for the analysis of data structured into units as formal abstractions of concepts of human thought [9]. The formal context defined by the FCA was considered for the dynamic corpus on chunking operation.

Considering a set of title tags $T$, a set of description tags $D$ and a set of incidence relations of context $I$ defined by the frequency of occurrence of the relation between $T$ and $D$ as depicted in (4): a set of formal contexts $C$ is defined by (3). Formal context was considered binding the use of RDF(S) constructs to remain in the first order logic.

The tag $T$ derived by the title was considered as a facet described by the tag $D$ derived by the description.

```
T: / Stocks| seesaw| in| volatile| trade|
D: | U.S.| stocks| slid| choppy| trading| dearth|
economic| news| left| investors| hesitant| make |
big| bests| amid| worries| potential| global|
recession
```

TABLE I. INCIDENCE MATRIX

| D \ T | stocks | seesaw | volatile | trade |
|---|---|---|---|---|
| U.S. | | | | |
| stocks | X | | | S |
| slid | | S | | |
| choppy | | S | S | |
| trading | S | | | X |
| dearth | | | | |
| economic | S | | | S |
| news | | | | |
| left | | | | |
| investors | S | | | S |
| hesitant | | | S | |
| make | | | | |
| big | | | | |
| bets | | S | | |
| worries | | | | |
| potential | | | | |
| global | | | | |
| recession | S | | | S |

$$(3) \quad C_n := (T_n, D_n, I_n)$$

On (3) the *set of incidence relations of context I* is defined by the matching between *T* and *D* tags by relation (4) allowing multiple associations among *D* tags and the faceted context defined by every *T* tag.

$$(4) \quad I \subseteq T \times D$$

Multiple matching was disambiguated by updating a Jaccard similarity coefficient (5) associated with the incidence relation of context $\forall i \in I$

Jaccard similarity coefficient, also known as the Tanimoto coefficient, measures the overlap of two sets. It is defined as the size of the intersection of the sets divided by the size of their union.

The Jaccard index is zero if two sets of the *incidence relation of context* are disjoint (i.e., they have no common members) and it is one if they are identical. Higher numbers indicate better agreement in the sets, so to disambiguate multiple matching the goal is to get as close to 0 as possible. [10]

$$(5) \quad J(T,D) = |T \cap D| / |T \cup D|$$

We consider a formal context as the association between an item in the title *T* and a set of terms in the description *D*.

It was built a matrix with the *T* and *D* sets of tags (Table1) where it is used:
- X marks to represent syntactically identical words;
- S marks to represent Similar words according to an OWL version of SUMO formal ontology (http://www.ontologyportal.org/) that has been mapped to all of the WordNet lexicon (http://wordnet.princeton.edu).

Considering any set of tags *T*, i.e. for A={stocks} the attribute on tags *D* derived are:
```
A'={stocks, trading, economic, investors, recession}
```

From those formal context can be derived:
```
(A,A')=({stocks}, {stocks, trading, economic,
        investors, recession })
(B,B')=({seesaw}, {slid, choppy, bets })
(C,C')=({volatile}, {choppy, hesitant })
(D,D')=({trade},{stocks, trading, economic,
        investors, recession })
```

Formal context are represented in a line diagram as a node:
```
(D, D') ------ (A, A')
```
The formal context (A, A') is called subcontext of (D,D'), and (D,D') is called supercontext of (A,A'). (A,A') is drawn below (D, D') and connected with a line.

Formal context can be combined, i.e. (F, F') is derived by (A, A'), (B, B') and (C, C')

```
(F, F')=({ stocks, seesaw, volatile
},{stocks, trading, economic, investors, recession,
slid, choppy, bets, hesitant })
```

By adding more formal context the diagram is extended step by step. The subcontext – supercontext relation defines an order on the set of all formal contexts.

For two contexts (A, A') and (F, F') this order is formalized as: (A, A') is smaller than (F, F') if A is a subset of F (and A' is a subset of F')

For each set of formal context exists always a unique greatest subcontext and a unique smallest supercontext.

FCA is not really suitable for real-time environments as it requires expensive computation, but a chunk semantic parsing is more efficient having been implemented with a finite state machine applied to very large text sources [3]. Finite state techniques have worst-case complexity O(n) in length of string.

We will discuss in the following sections a definition of a partial order to hierarchically organize the related data, such as a specific statistic/semantic relationship.

*C. Chunks proposed as Tags to the user*

A selected number of chunks, defined according to the *Formal Context* (*C*) as seen in *IV.B* (*Incidence of context on folksonomy technology*), are proposed to the user as folksonomy tags for the correlated uri *Resource* (*R*).

The defined *Formal Context* (*C*) is used in the *Folksodriven tags* defined in (1) where it is correlated to the uri *Resource* (*R*) from where *C* is derived and the *Time Exposition* (*E*), that is defined by the clickthrough rate by the users that have chosen that particular *Formal Contest* (*C*) as folksonomy tag for the uri *Resource* (*R*).

We will discuss in the following sections a definition of a partial order to hierarchically organize the related data, such as a specific statistic/semantic relationship.

*D. Folksodriven as a Network*

On scientific literature folksonomies are modeled with patterns and mathematical rules of networks [11, 12]. The most significant contribution of network theory is the demonstration of the "mathematical power law complex systems" as well as the existence of stability in the tags used on a website over time [13].

The theoretical starting point considered is the emerging scientific network theory that aims to demonstrate mathematically that all complex networks follow a number of general regularities. The connection between the different vertices in the network forms a pattern. The pattern of all complex systems is the power law that arises because the network expands with the addition of new vertices which must be connected to other vertices in the network.

A network with a degree distribution that follows a power law is a scale-free network. An important characteristic of scale-free networks is the clustering coefficient distribution, which decreases as the node degree increases following a power law [11, 13]. We consider a Folksodriven network in which nodes are Folksodriven tags and links are semantic acquaintance relationships between them according to the SUMO formal ontology (http://www.ontologyportal.org) that has been mapped to the entire WordNet lexicon (http://wordnet.princeton.edu). It is easy to see that Folksodriven tags tend to form small groups in which tags are close related to each one, so we can think of such groups as a complete graph. In addition, the Folksodriven tags of a group also have a few acquaintance relationships to Folksodriven tags outside that group. Some Folksodriven tags, however, are so related to other tags (e.g., workers, engineers) that are connected to a large number of groups. Those Folksodriven tags may be considered the hubs responsible for making such network a scale-free network [13, 14].

In a scale-free network most nodes of a graph are not neighbors of one another but can be reached from every other by a small number of hops or steps, considering the mutual acquaintance of Folksodriven tags.

V. CLUSTERING ANALYSIS

A network can be represented by an adjacency matrix, A=[aij], that encodes whether/how a pair of nodes is connected. For weighted networks, the adjacency matrix A reports the connection strength between node pairs. A is a symmetric matrix and for our convention we consider the diagonal elements of the matrix A as 1. A node is defined for the same uri resource *r* in the vector space by the ternary relation $X = C \times E \times R$. A node *i* of a network is connected to other nodes called the neighbors of node *i* and define the closest neighborhood of this node.

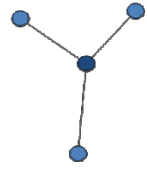 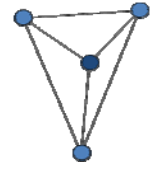

Figure 1: Clustering Coef. of the black central node = 0

Figure 2: Clustering Coef. of the black central node = 1

We consider a network that is weighted by the *Time Exposition* (*E*). So according to the Folksodriven notation (1) we depict the *Formal Context* (*C*) of different *Resources* (*R*) that are weighted by the *Time Exposition* (*E*).

For weighted networks we consider the connectivity as the sum of connection strengths to other nodes (row sum of the adjacency matrix).

$$(6) \quad Connectivity_i = k_i = \sum_{j \neq i} a_{ij}$$

To estimate the clustering coefficient for a node we then find the number of distinct links between these neighbors. Clustering Coefficient measures the cliquishness of a particular node: « A node is cliquish if its neighbors know each other ». Generalizing directly to weighted networks from [14] we consider a Clustering Coefficient as a local property measuring the connectedness of the neighbours of the vertex. The definition of the clustering coefficient in classical networks is illustrated by Watts [15].

$$(7) \quad ClusterCoef_i = \frac{\sum_{l \neq i} \sum_{m \neq i,l} a_{il} a_{lm} a_{mi}}{\left(\sum_{l \neq i} a_{il}\right)^2 - \sum_{l \neq i} a_{il}^2}$$

We use the Clustering Coefficient to find "natural" grouping of Folksodriven tag relations.

If the clustering method is exclusive, it means that each object can be assigned to only one cluster at a time allowing a minimum connectedness of the neighbors of the vertex referred to the uri resource *r*.

Overlapping strategy allows multiple assignments of any object allowing a maximum connectedness of the neighbors of the vertex referred to the uri resource *r*.

The clustering coefficient relates to the local "cliqueness" and the higher it is the better the network can withstand the effect of link removal, which tends to fragment the network thereby making it less stable (more random).

The *Global Clustering Coefficient K* is the average over all nodes for the same uri resource $r$: $K=<Kr(i)>r$.
Considering different kind of set according to other uri resources than $r$ it's possible to see the variation of the Folksodriven Clustering Coefficient for the ternary relation:
$X = C \times E \times R$

## VI. EXPERIMENTS

### A. Setup

Throughout realizing a test network model in a simulated environment we compute clustering coefficients (to determine the neighbor connectivity), for a scan over the relevant parameter space.

The original Folksodriven network was compared with two kinds of different graphs developed from the same original data:
• Scale-free Random graphs: the process to generate graphs adds vertices one at a time joining to a fixed number of starting vertices that are chosen with probability proportional to the graph degree. This process was presented as a simple model of the growth of the world-wide-web by Barabasi and Albert [14]. The vertices are created by picking the three endpoints *(C, E, R)* from the observed data set and the relative Folksodriven assignments |*X*|. Each starting vertex chosen is chosen with probability proportional to its degree.
• Scale-free Diffeomorphic graphs: the process to generate graphs uses an invertible function that maps one dimension to another, such that both the function and its inverse have derivatives of all orders being smooth. The vertices are created by picking the three endpoints *(C, E, R)* from the observed data set and rearranging each dimension of *X* mapping it to another dimension [15]. The created graphs have the same degree as the original one.

All data were obtained from averages over 100 identical network realizations with a sample of 400 nodes taken randomly from each graph and computing the Clustering Coefficient from each of those vertex to all others in the Folksodriven tags using depth-first search. Twenty runs were performed to ensure consistency for all experiments involving randomness and diffeomorphic. The reason for these relatively modest network sizes is that, at every time-step, all network distances have to be evaluated.

In order to check whether our observed Folksodriven graph exhibits scale-free characteristics, we compared the characteristic Clustering Coefficients with Random graphs and Diffeomorphic graphs of a size equal in all dimensions *C*, *E*, and *R* as well as *X* to the respective Folksodriven under consideration (same number of nodes and links). We now turn to the clustering and neighbor connectivity of the emerging networks. For larger time exposition *E* the Clustering Coefficients become drastically smaller, as expected for the *E* that tends to infinitive limit and *C* that tends to zero limit, as we can see in Figure 4.

### B. Data set analysis

Figure 3 shows the results for the Clustering Coefficients for the dataset, plotted against the starting Folksodriven tags |*X*| assigned. In the Folksodriven dataset (Figures 3 and 4), it can be seen that both clustering coefficients are higher than those for the Diffeomorphic and Random graphs. This could be an indication of coherence in the tagging behavior. For example: when a given set of tags is attached to a certain kind of resources, users do so consistently.

On the other hand, the characteristic Folksodriven Clustering Coefficient (Figure 3) is considerably smaller than for the scale-free Random graphs, though not as small as for the Diffeomorphic graphs. The Clustering Coefficient has remained almost constant at about 3. While the number of nodes has grown about twenty times during the observation period. This implies that on average, every *Formal Context* (*C*), *Time Exposition* (*T*) and *Resource* (*R*) defined on the original data set can be reached within 3 mouse clicks from any given page. We can argue that even if the folksonomy grows to millions of nodes, everything in it is still reachable within few hyperlinks. This attest the context of "serendipitous discovery" [17] of contents in the folksonomy community.

## VII. RELATED WORKS

### A. Folksonomies

There are a lot of works on collaborative tagging and folksonomies that can be discussed in related works. More cognizances can be given to the now extensive folksonomy literature and how the context and use has developed over the past five years or so. To name a few:
• A general overview of folksonomies structure and their dynamics is provided by [16, 17].
• Particular aspects of folksonomies have been elaborated in more detail, as ranking of contents, discovering trends in the tagging behavior of users [11], or learning taxonomic relations from tags [21, 22, 23].
• Semantic similarity measures among users based solely on their annotation metadata is observable among users who lie close to each other in the social network [24].
• In collaborative tagging system for tasks like synonym detection and discovery of context hierarchies many researchers introduced measures of tag similarity that are investigated on [25] according to the context of a given semantic application.
• Regularities are discovered on [26] considering collaborative tagging activity by users to propose a dynamical model of collaborative tagging that predicts these stable patterns and relates them to imitation and shared knowledge.
• On [27] is analyzed the potential impact of the individual and social behavior on the design of mechanisms that support tagging systems

### B. Networks and graph theories

The theoretic notions on Folksodriven graph depicted on Section *IV.D* are derived using contexts from social network analysis, graph theory [18, 19, 20, 21], as well as statistical scale-free networks on the web and online communities, and it is widespread used in many different fields [11, 13, 13]. Networks related to folksonomy, as with other different human based social networks, have peculiar characteristics, the tailed distribution of a complex interaction between human agents is described in reference [15]. Watts and Strogatz introduced the notion of clustering coefficient [30].

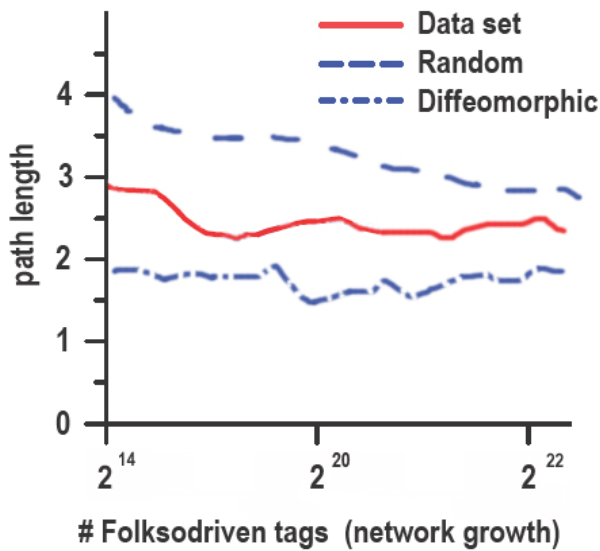
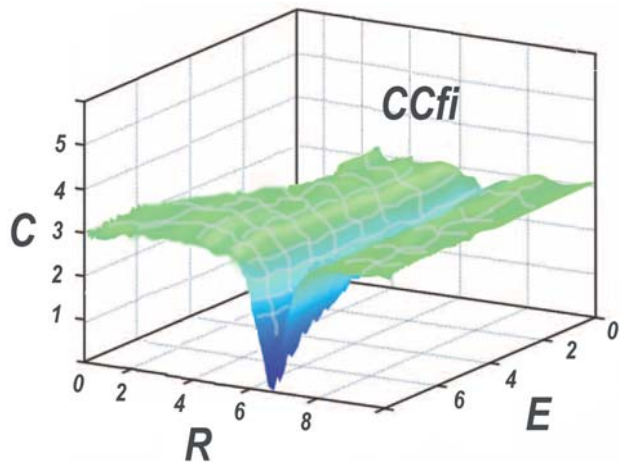

Figure 3: Folksodriven set of data compared with the corresponding Random and Diffeomorphic graphs. The measure is repeated during the network growth as a function of the number of events determined by folksodriven tags. It is depicted how the characteristic *path length* takes quite similar values for the corresponding Random and Diffeomorphic graph.

Figure 4: It is depicted the *Clustering Coefficient* in the space delimited by the *Formal Context* (C), the *Time Exposition (E)* and the *Resource (R)*. For larger *Time Exposition (E)* the clustering coefficients become drastically smaller, as expected for the *E* that tends to infinitive limit and *C* that tends to zero limit.

### VIII. OVERVIEW AND CONCLUSION

In this paper, we proposed and analyzed the network structure of the Folksodriven tags thought as folsksonomy tags suggestions for the user on a dataset built on chosen websites. Folksodriven are driven by the click-through rate (CTR) thought as time exposition of the suggested tags and the Formal Concept Analysis (FCA). We observed that the hypergraphs of the Folksodriven are highly connected and that the relative path lengths are relatively low, facilitating thus the "serendipitous discovery" of interesting contents for the users.

### IX. FUTURE WORK

Section *IV.D* (*Folksodriven as a Network*) suggests that the Folksodriven consists of densely-connected communities. A line of research that will benefit from the observations proposed in this paper is the identification of Communities [28].

Folksodriven notation can be used in a community environment to make explicit groups of interests which already exist intrinsically in a Folksodriven system, to be used for recommender systems, opinion mining, and sentiment analysis. Chunking was used in this work as a starting point but it is at a very low semantic level.

News articles, blog posts and e-mails often lack a systematic reference list that could be used to make a reference [29]. Yet they, too, are part of what makes an idea influential. This allows important shifts in terminology to be tracked down to their origins, which offers a way to identify truly ground-breaking work [31], the sort of stuff that introduces new contexts so an user contribution can be determined by looking at how big a shift it creates in the structure of the Folksodriven notation.

Another line of research that we are currently pursuing is the dynamic taxonomies/faceted search and on-the-fly clustering (a la Vivisimo) that would support a much more effective interaction with the user (http://vivisimo.com).

**Massimiliano Dal Mas** is an engineer at the Web Services division of the Telecom Italia Group, Italy. His interests include: user interfaces and visualization for information retrieval, automated Web interface evaluation and text analysis, empirical computational linguistics, and text data mining. He received BA, MS degrees in Computer Science Engineering from the Politecnico di Milano, Italy. He won the thirteenth edition 2008 of the CEI Award for the best degree thesis with a dissertation on "Semantic technologies for industrial purposes" (Supervisor Prof. M. Colombetti).

This paper has been accepted to the International Conference on Social Computing and its Applications (SCA 2011)

Sydney Australia, 12-14 December 2011

SCA2011 Accepted Papers:

- **Cluster Analysis for a Scale-Free Folksodriven Structure Network**
  *Massimiliano Dal Mas*

Dear Massimiliano,

Congratulations on your paper acceptance to SCA2011 - Your paper ID is SCA#62. Your paper is accepted as a regular one titled Cluster Analysis for a Scale-Free Folksodriven Structure Network.

Thanks for your attention.

We look forward to seeing you in Sydney Australia during 12-14 December 2011.

SCA2011 organising committee